\documentclass[a4paper,11pt,aps,reprint,showkeys,showpacs]{revtex4-1}

\usepackage[utf8]{inputenc}
\usepackage[T1]{fontenc}
\usepackage{amsmath,amssymb,amsfonts,amsthm}
\usepackage{graphicx}
\usepackage{hyperref}

\newcommand{\xv}{\mathbf{x}}
\newcommand{\pv}{\mathbf{p}}
\newcommand{\iu}{\mathbf{i}}
\newcommand{\trho}{\tilde{\rho}}
\newcommand{\tpsi}{\tilde{\psi}}
\newcommand*\diff{\mathop{}\!\mathrm{d}}

\begin{document}

\title{On a Conjecture Regarding Fisher Information}%

\author{G. Bellomo}%
\email{gbellomo@fisica.unlp.edu.ar}%
    \affiliation{Instituto de Física La Plata (IFLP-CONICET),
        and Departamento de Física,\\ Facultad de Ciencias Exactas,
        Universidad Nacional de La Plata,\\
        115 and 49, C.C. 67, 1900 La Plata, Argentina}%

\author{A.R. Plastino}%
\email{arplastino@unnoba.edu.ar}
    \affiliation{CeBio y Secretaria de Investigacion,\\
        Universidad Nacional del Noroeste de la Prov. de Buenos Aires
        - UNNOBA and CONICET,\\
        R. Saenz Pena 456, Junin, Argentina}%

\author{A. Plastino}
\email{angeloplastino@gmail.com}%
    \affiliation{Instituto de Física La Plata (IFLP-CONICET),
        and Departamento de Física,\\ Facultad de Ciencias Exactas,
        Universidad Nacional de La Plata,\\
        115 and 49, C.C. 67, 1900 La Plata, Argentina}%

\begin{abstract}
Fisher's information measure $I$ plays a very important role in
diverse areas of theoretical physics. The associated measures
$I_x$ and $I_p$, as functionals of quantum probability distributions
defined in, respectively, coordinate and momentum spaces, are the
protagonists of our present considerations. The product ${I_xI_p}$
has been conjectured to exhibit a non trivial lower bound in
{\textit{[Phys. Rev. A \textbf{62} 012107 (2000)]}}. More explicitly, this conjecture says that for any pure state of a particle in one dimension ${I_xI_p\geq4}$. We show here
that such is not the case. This is illustrated, in particular, for
pure states that are solutions to the free-particle Schr\"odinger
equation. In fact, we construct a family of counterexamples to the conjecture, corresponding to time-dependent solutions of the free-particle Schr\"odinger equation. We also conjecture that any normalizable time-dependent solution of this equation verifies ${I_xI_p\rightarrow0}$ for ${t\rightarrow\infty}$.
\end{abstract}


\maketitle

\section{Introduction}
A very important information measure, with manifold physical
applications, was conceived by R.A.~Fisher in the 1920's --for
detailed discussions see~\cite{Fried98,Fried04,Carr10,Penn04}--.
Recent developments show that Fisher’s information has a fundamental
role in quantum mechanics~\cite{Regi99,Nagy10,Sen07,Luo02,
Facc10,Giro13,Choi11,Liu14,Nich14,SMPD11,ToDe14,DPSV12,DEPS11,FPPS02,PPCG00}.
In particular, it allows for the formulation of new quantum
uncertainty principles~\cite{Luo00,Rome05,Sanc06,Dehe07,Toth13}.
It is usually abbreviated as $I$ and can be thought of as a
measure of the expected error in a measurement~\cite{Fried98}.

A particular instance of great relevance is that of translational
families~\cite{Fried98}. These are distribution functions whose
form remains invariant under displacements of a shift parameter
$\theta$. Thus, they are shift invariant distributions (in a Mach
sense, there is no absolute origin for $\theta$). The measure
exhibits Galilean invariance~\cite{Fried98}. Given a probability density $f(\xv,\theta)$, with
$\xv\in\mathbb{R}^D$ and $\theta=(\theta_i)_{1\leq i\leq n}$ a
family of parameters, the concomitant Fisher matrix is~\cite{Fried95}
    \begin{equation} \label{eq:fisher_def}
    I_{jk} := \int{\frac{1}{f(\xv,\theta)}
    \left(\frac{\partial f}{\partial\theta_j}\right)
    \left(\frac{\partial f}{\partial\theta_k}\right) \diff{\xv}},
    \end{equation}
where $\diff{\xv}=\prod_{k=1}^D{\diff{x_k}}$ is the volume element
in $\mathbb{R}^D$. In particular, for $\theta\in\mathbb{R}^D$,
one defines translational families $f(\xv-\theta)$, with elements
${I_{jk} = \int{\frac{1}{f} (\partial_j f) (\partial_k f)
\diff{\xv}}}$, where $\partial_i$ represents the partial
derivative with respect to the coordinate $x_i$. The trace of this matrix, given by ${I = \int{\frac{1}{f}
[\sum_{k=1}^D (\partial_k f)^2] \diff{\xv}}}$, is a good
uncertainty indicator for probability distributions associated to
quantum wave functions \cite{Hall00}. If $\psi(\xv)$ is a
normalized wave function in coordinate space ($D$-dimensions) and
 ${\tilde{\psi}(\pv) = (2\pi)^{-D/2} \int{e^{-\iu\xv\cdot\pv}
\psi(\xv) \diff{\xv}}}$ is its momentum-counterpart, the
corresponding probability densities are, respectively,
${\rho(\xv) = |\psi(\xv)|^2}$ and ${\trho(\pv) =
|\tilde{\psi}(\pv)|^2}$, with associated Fisher measures
    \begin{align} \label{eq:psi_fisher}
    I_\xv = \int{\frac{1}{\rho}[\nabla_\xv \rho]^2 \diff\xv} \,, \\
    I_\pv = \int{\frac{1}{\trho}[\nabla_\pv \trho]^2 \diff\pv},
    \end{align}
allow one to study uncertainty relations via the product
${I_\xv I_\pv}$~\cite{Hall00}.

For instance, one can demonstrate that, if $\psi(\xv)$ (or
$\tilde{\psi}(\pv)$) is real, then ${I_\xv I_\pv \geq
4D^2}$~\cite{SMPD11}, with equality for coherent states of the
harmonic oscillator (HO)~\cite{Hall00}. For general, mixed, states
it is clear that the product ${I_\xv I_\pv}$ does not possess
a non trivial lower bound. (For example, one can use thermal HO
states, represented by Gaussian distributions in both coordinates
and momenta, in the high temperature limit.) In the case of pure
states, though, the existence of such a lower bound for
${I_\xv I_\pv}$ was an open question. Hall conjectured that the
relation ${I_x I_p}\geq4$ might hold in general for pure states
in one dimension~\cite[p.~3]{Hall00}. We will next present a couple
of counterexamples that show this conjecture to be incorrect.
Our examples give rise to a new conjecture: for a bounded wave
function one has ${I_\xv I_\pv \rightarrow 0}$ when
${t\rightarrow\infty}$.

\section{Counterexamples}

Our first example is taken from the considerations (in a different
context) of reference~\cite{SMPD11}. A free-particle's one
dimensional wave packet $\psi(x,t)$ (unit mass) evolves according
to Schr\"odinger's equation
    \begin{equation} \label{eq:schr}
    \iu \frac{\partial\psi}{\partial t} =
        -\frac{1}{2}\frac{\partial^2\psi}{\partial x^2} \,.
    \end{equation}
Setting the initial conditions
    \begin{equation} \label{eq:exA_1}
    \begin{aligned}
    &\psi(x,0) = A_0 \exp{\left[-\frac{x^2}{2\Delta^2}\right]} \,, \\
    &\tpsi(p,0) = \tilde{A}_0 \exp{\left[-\frac{\Delta^2 p^2}{2}\right]} \,,
    \end{aligned}
    \end{equation}
with ${A_0=\Delta^{-1/2}\pi^{-1/4}}$,
${\tilde{A}_0=\Delta^{1/2}\pi^{-1/4}}$ and $\Delta>0$, that
correspond to a Gaussian packet, one finds the solution
    \begin{equation} \label{eq:exA_2}
    \psi(x,t) = A(t)
    \exp{\left[-\frac{x^2}{2\Delta^2(1+\iu t/\Delta^2)}\right]} \,,
    \end{equation}
where $A(t)=A_0(1+\iu t/\Delta^2)^{-1/2}$. The associated probability
densities are
    \begin{equation} \label{eq:exA_3}
    \begin{aligned}
    &\rho(x,t) = \frac{\Delta}{\sqrt{\pi(\Delta^4+t^2)}}
    \exp{\left[-\frac{\Delta^2 x^2}{\Delta^4+t^2}\right]} \,\\
    &\trho(p,t) = \frac{\Delta}{\sqrt{\pi}}\exp[-\Delta^2 p^2] \,.
    \end{aligned}
    \end{equation}
The product ${I_xI_p=4\Delta^4(\Delta^4+t^2)^{-1}}$ obeys the relation
$I_xI_p<4$ for $t>0$. Also, one has ${I_xI_p\rightarrow0}$ for
$t\rightarrow\infty$.

\vspace{4mm}

We pass now to another free-particle solution, given by the first partial derivative of $\psi(x,t)$
with respect to $x$: ${\psi^{(1)}(x,t)\propto\partial_x\psi(x,t)}$ [see
Eq.~\eqref{eq:schr}],
correctly normalized. It is easy to see that $\psi^{(1)}(x,t)$
is a solution by deriving both members of Eq.~\eqref{eq:schr},
i.e.
    \begin{equation} \label{exB_1}
    \iu \frac{\partial}{\partial t}\frac{\partial \psi}{\partial x}
     = -\frac{1}{2}\frac{\partial^2}{\partial x^2}
        \frac{\partial \psi}{\partial x} \,.
    \end{equation}
The new solution is
    \begin{equation} \label{eq:exB_2}
    \psi^{(1)}(x,t) = A^{(1)}(t)
    \exp{\left[-\frac{x^2}{2\Delta^2(1+\iu t/\Delta^2)}\right]} \,,
    \end{equation}
with  ${A^{(1)}(t)=-2^{1/2}\pi^{1/4}\Delta^{3/2}(\Delta^2+\iu t)^{-3/2}}$.
The two corresponding densities are
    \begin{equation} \label{eq:exB_3}
    \begin{aligned}
    &\rho^{(1)}(x,t) = \frac{2 \Delta^3}{\sqrt{\pi \left(\Delta^4+t^2\right)^3}}
        x^2 \exp{\left[-\frac{\Delta^2 x^2}{\Delta^4+t^2}\right]} \,, \\
    &\trho^{(1)}(p,t) = \frac{2 \Delta^3}{\sqrt{\pi}} p^2
        \exp{[-\Delta ^2 p^2]} \,.
    \end{aligned}
    \end{equation}
The product is ${I^{(1)}_xI^{(1)}_p=36\Delta^4(\Delta^4+t^2)^{-1}}$, verifying
${I^{(1)}_xI^{(1)}_p<4}$ for $t>2\sqrt{2}\Delta^2$, and ${I^{(1)}_xI^{(1)}_p\rightarrow0}$
when ${t\rightarrow\infty}$.

\vspace{4mm}

In general, one can show that the whole family of solutions of
Eq.~\eqref{eq:schr} given by successive derivatives of $\psi(x,t)$,
i.e., the set
${\{\psi^{(n)}(x,t)|\psi^{(n)}(x,t)=N_n\partial^n_x\psi(x,t),n=0,1,2...\}}$,
 verifies that ${I_xI_p\rightarrow0}$ when $t\rightarrow\infty$,
with $N_n$ the pertinent normalization constants. Thus, the
family ${\{\psi^{(n)}(x,t)\}_{n\in\mathbb{N}_0}}$ yields infinite
counterexamples to Hall's conjecture. To see this, one needs
first to rewrite the Fisher measure in wave function's terms, so
that Eq.~\eqref{eq:psi_fisher} becomes equivalent to
    \begin{equation} \label{eq:fisher_w}
    I_{\xv} = 4\int{(\nabla_\xv|\psi|)^2\diff{\xv}} \,,
    \end{equation} or, in one dimension,
${I_x=\int{(\partial_x|\psi|)^2\diff{x}}}$. Further,
${|\psi|=\psi^*\psi}$. Thus, $I_x$ can be expressed in terms of
$\psi$ and $\psi^{(1)}$. In one dimension one has
    \begin{equation} \label{eq:fisher_w2}
    I_x = 4\int{(\psi^{(1)*}\psi+\psi^*\psi^{(1)})^2\diff{x}} \,.
    \end{equation}
In general, for  $\psi^{(k)}$, the Fisher's measure associated to
the distribution $|\psi^{(k)}|^2$ becomes
    \begin{equation} \label{eq:fisher_wn}
    I_x^{(k)} = 4\int{(\psi^{(k+1)*}\psi^{(k)}
        + \psi^{(k)*}\psi^{(k+1)})^2\diff{x}} \,.
    \end{equation}

We show now that the integrand tends to zero for $t\rightarrow\infty$.
Thus, ${I^{(k)}_x\rightarrow0}$ in such a limit. Actually, we will
show that ${\psi^{(k)}(x,t)\rightarrow0}$ for ${t\rightarrow\infty}$.
The $k$-th derivative of ${\psi(x,t)\equiv\psi^{(0)}(x,t)}$ is
proportional to the $k$-th derivative of a Gaussian distribution, given by
    \begin{align} \label{eq:gauss_herm}
    \psi^{(k)}(x,t) &= N_k(t) \frac{\partial^k}{\partial x^k}\psi^{(0)}(x,t) \notag\\
    &= N_k(t) \frac{\partial^k}{\partial x^k} \left(A(t) e^{-c(t)^2 x^2}\right) \\
    &= N_k(t) A(t) (-1)^k c(t)^k H_k(c(t)x) e^{-c(t)^2 x^2} \notag\\
    &= N_k(t) (-1)^k c(t)^k H_k(c(t)x) \psi^{(0)}(x,t) \notag\,
    \end{align}
where ${c(t)^2=(2(\Delta^2+\iu t))^{-1}}$ and $H_k(y)$ is the
Hermite polynomial of degree $k$ in the variable $y$. The time-dependent quantities $c(t)$,
$\psi^{(0)}(x,t)$, and $A(t)$ vanish for ${t\rightarrow\infty}$.
What is the behavior of $N_k$? Let us see what happens with
$\tpsi^{(k)}(p,t)$, the $k$-th solution in momentum space,
corresponding to the Fourier transform of $\psi^{(k)}(x,t)$. We
have
    \begin{equation} \label{eq:mom_k}
    \begin{aligned}
    \tpsi^{(k)}(p,t) &= \frac{1}{\sqrt{2\pi}}\int{e^{-\iu xp}\psi^{(k)}(x,t)\diff x} \\
        &= \frac{N_k(t) A(t)}{\sqrt{2\pi}}\int{e^{-\iu xp} \frac{\partial^k}{\partial x^k} e^{-c(t)^2 x^2} \diff x} \\
        &= \frac{N_k(t) A(t) (\iu p)^k}{\sqrt{2c(t)^2}}\exp\left[-\frac{p^2}{4c(t)^2}\right] \\
        &= N_k(t) (\iu p)^k \tpsi^{(0)}(p,t).
    \end{aligned}
    \end{equation}
Demanding normalization leads to
    \begin{equation} \label{eq:mom_norm}
    \begin{aligned}
    1 &= \int{\tpsi^{(k)}(p,t)\tpsi^{(k)*}(p,t) \diff{p}} \\
        &= |N_k(t)|^2 \int{p^{2k} |\tpsi^{(0)}|^2 \diff{p}} \\
        &= |N_k(t)|^2 \frac{\Delta}{\sqrt{\pi}} \Gamma(k+\tfrac{1}{2}) \Delta^{-2k-1} \,.
    \end{aligned}
    \end{equation}
Thus,
    \begin{equation} \label{eq:norm}
    |N_k(t)|^2 = \frac{\sqrt{\pi}\Delta^{2k}}{\Gamma(k+\tfrac{1}{2})} \,,
    \end{equation}
and ${|N_k(t)|^2}$ is time-independent. We can conclude that ${|\psi^{(k)}(x,t)|\rightarrow0\;\forall\,k}$ (see Eq.~\eqref{eq:psi_lim} below). Now, remind that
    \begin{equation} \label{eq:psi0}
    \psi^{(0)}(x,t)=A(t)e^{-c(t)^2x^2} \,,
    \end{equation}
with ${A(t)=\pi^{-1/4}\sqrt{2\Delta}c(t)}$, and
    \begin{equation} \label{eq:tpsi0}
    \tpsi^{(0)}(p,t)=\frac{\sqrt{\Delta}}{\pi^{1/4}}\exp\left[-\frac{p^2}{4c(t)^2}\right].
    \end{equation}
One finds the following limits for the absolute values of the wave
functions:
    \begin{equation} \label{eq:psi_lim}
    \begin{aligned}
    |\psi^{(k)}&(x,t)|^2 = |N_k(t) c(t)^k H_k(c(t)x) \psi^{(0)}(x,t)|^2 \\
            &\sim |c(t)|^{2k} |A(t)|^2 e^{-2\Re(c(t)^2)x^2} \\
            &= \frac{\Delta}{(\Delta^4+t^2)^{\tfrac{k+1}{2}}} \exp\left[-\frac{\Delta^2x^2}{\Delta^4+t^2}\right] \xrightarrow[t\rightarrow\infty]{} 0 \,,
    \end{aligned}
    \end{equation}
    \begin{equation} \label{eq:tpsi_lim}
    \begin{aligned}
    |\tpsi^{(k)}(p,t)|^2 &= |N_k(t) (\iu p)^k \tpsi^{(0)}(p,t)|^2 \\
            &= \frac{\Delta^{2k+1}}{\Gamma(k+\tfrac{1}{2})} p^{2k} e^{-\Delta^2 p^2} \,.
    \end{aligned}
    \end{equation}
Eq.~\eqref{eq:psi_lim} indicates that all functions
${\psi^{(k)}(x,t)}$ vanish for ${t\rightarrow\infty}$. Accordingly,
from Eq.~\eqref{eq:fisher_wn} we find ${I_x^{(k)}\rightarrow0}$
for ${t\rightarrow\infty}$ for all ${k=0,1,2...}$. Further,
Eq.~\eqref{eq:tpsi_lim} shows that ${\tpsi^{(k)}(p,t)}$ does not
vanish in this limit. In fact, $|\tpsi^{(k)}(p,t)|$ does not
depend on $t$.

So as to understand what happens with $I_p^{(k)}$ let us see
an expression analogous to Eq.~\eqref{eq:fisher_wn} in momentum
space:
    \begin{equation} \label{eq:fish_pn}
    I_p^{(k)} = 4\int{(\tpsi^{(k+1)*}\tpsi^{(k)}+\tpsi^{(k)*}\tpsi^{(k+1)})^2\diff{p}} \,,
    \end{equation}
Expanding the integrand using Eq.~\eqref{eq:mom_k} we have
    \begin{equation} \label{eq:pn_integ}
    \begin{aligned}
    \tpsi^{(k+1)*}&\tpsi^{(k)} + \tpsi^{(k)*}\tpsi^{(k+1)} \\
    &= 2\Im(N_kN_{k+1}^*) p^{2k+1} |\tpsi^{(0)}(p,t)|^2 \,.
    \end{aligned}
    \end{equation}
Introducing this into Eq.~\eqref{eq:fish_pn}, and remembering
that both $N_k$ and $c(t)$ are independent of $p$, we find
    \begin{equation} \label{eq:fish_pk}
    \begin{aligned}
    I_p^{(k)} &= 16 \Im(N_kN_{k+1}^*)^2 \frac{\Delta^2}{\pi} \int{p^{4k+2} e^{-2\Delta^2p^2}} \\
        &= 16 \Im(N_kN_{k+1}^*)^2 \frac{\Delta^2}{\pi} \frac{\Gamma(2k+\frac{3}{2})}{(2\Delta^2)^{2k+3/2}} \,.
    \end{aligned}
    \end{equation}
Since ${|N_k(t)|^2\sim1}$ (see Eq.~\eqref{eq:norm}),
one has ${|N_k(t)N_{k+1}|^2\sim1}$ and thus $I_p^{(k)}$ becomes
bounded. Thus, there exists ${I_{p,\max}^{(k)}\in\mathbb{R}_{>0}}$
such that ${I_p^{(k)}\leq I_{p,\max}^{(k)}}$ for all
$k=0,1,2...$. We conclude that ${I_x^{(k)}I_p^{(k)}\rightarrow0}$
for $t\rightarrow\infty$ for the whole family of solutions
${\{\psi^{(k)}(x,t)\}_{k\in\mathbb{N}_0}}$.

\section{Conclusions}
We conclude by reiterating that we have found an infinite number
of counterexamples to the conjecture ${I_x\,I_p \ge 4}$, for pure
states, put forward in~\cite{Hall00}. On the basis of these results, we conjecture that, for any normalizable wave function $\psi(x,0)$, the corresponding time-dependent solution $\psi(x,t)$ of the free-particle Schr\"odinger equation, satisfies ${I_xI_p\rightarrow0}$ for ${t\rightarrow\infty}$.

\section*{\scriptsize Acknowledgement}
The authors would like to thank
Pablo Sanchez Moreno for
useful discussions.


\end{document}